\documentclass[12pt,preprint]{aastex}

\newcommand{\msec}[2]{$#1\mbox{$''\mskip-7.6mu.\,$}#2$}
\newcommand{\mmsec}[2]{$#1\mbox{$^s\mskip-7.6mu.\,$}#2$}

\newcommand{\Lsun}{L$_{\odot}$}
\newcommand{\Msun}{M$_{\odot}$}
\newcommand{\Rsun}{R$_{\odot}$}

\newcommand{\sbeamp}[5]{{$#1\mbox{$''\mskip-7.6mu.\,$}#2$} $\times$ {$#3\mbox{$''\mskip-7.6mu.\,$}#4$}; P.A.=$+#5^\circ$}
\newcommand{\sbeamm}[5]{{$#1\mbox{$''\mskip-7.6mu.\,$}#2$} $\times$ {$#3\mbox{$''\mskip-7.6mu.\,$}#4$}; P.A.=$-#5^\circ$}

\begin{document}

\title{Confirmation of a recent bipolar ejection \\ 
in the very young hierarchical multiple system IRAS~16293--2422}

\author{Gerardo Pech\altaffilmark{1} and Laurent Loinard}
\affil{Centro de Radiostronom\'{\i}a y Astrof\'{\i}sica, 
       Universidad Nacional Aut\'onoma de M\'exico,\\
       Apartado Postal 72--3 (Xangari), 58089 Morelia, Michoac\'an, M\'exico;\\       g.pech,l.loinard@crya.unam.mx}

\altaffiltext{1}{Also at: Facultad de Ingenier\'{\i}a, Universidad
  Aut\'onoma de Yucat\'an, Apartado Postal 150 (Cordemex), M\'erida,
  Yucat\'an, M\'exico.}

\author{Claire J.\ Chandler}
\affil{National Radio Astronomy Observatory, P.O.\ Box O, Socorro, NM 87801; cchandle@nrao.edu}

\author{Luis F.\ Rodr\'{\i}guez and Paola D'Alessio}
\affil{Centro de Radiostronom\'{\i}a y Astrof\'{\i}sica, 
       Universidad Nacional Aut\'onoma de M\'exico,\\
       Apartado Postal 72--3 (Xangari), 58089 Morelia, Michoac\'an, M\'exico;\\       l.rodriguez, p.dalessio@crya.unam.mx}

\author{Crystal L.\ Brogan}
\affil{National Radio Astronomy Observatory, 520 Edgemont Road, \\
Charlottesville, VA 22903-2475; cbrogan@nrao.edu}

\author{David J.\ Wilner and  Paul T.P.\ Ho\altaffilmark{2}}
\affil{Harvard-Smithsonian Center for Astrophysics, 60 Garden Street, 
       Cambridge, MA 02138;\\ dwilner, pho@cfa.harvard.edu}

\altaffiltext{2}{Also at: Academia
Sinica, Institute of Astronomy and Astrophysics, Taipei 106, Taiwan}

\begin{abstract} 

  We present and analyze two new high-resolution
  ($\sim$ \msec{0}{3}), high-sensitivity ($\sim$ 50 $\mu$Jy
  beam$^{-1}$) Very Large Array 3.6 cm observations of IRAS~16293--2422
  obtained in 2007 August and 2008 December. The 
  components A2$\alpha$ and A2$\beta$
  recently detected in this system are still present, and have moved roughly 
  symmetrically away from source A2 at a projected velocity of 30--80
  km s$^{-1}$. This confirms that A2$\alpha$ and A2$\beta$ were formed
  as a consequence of a very recent bipolar ejection from
  A2. Powerful bipolar ejections have long been known to occur in
  low-mass young stars, but this is --to our knowledge-- the first
  time that such a dramatic one is observed from its very beginning.
  Under the reasonable assumption that the flux detected at radio
  wavelengths is optically thin free-free emission, one can estimate the
  mass of each ejecta to be of the order of 10$^{-8}$ \Msun. If the 
  ejecta were created as a consequence of an episode of enhanced
  mass loss accompanied by an increase in accretion onto the protostar,
  then the total luminosity of IRAS~16293--2422 ought to have increased 
  by 10--60\% over the course of at least several months. 
  Between A2$\alpha$ and A2$\beta$, component A2 has reappeared, and
  the relative position angle between A2 and A1 is found to have
  increased significantly since 2003--2005. This strongly suggests
  that A1 is a protostar rather than a shock feature, and that the
  A1/A2 pair is a tight binary system. Including component B,
  IRAS~16293--2422 therefore appears to be a very young hierarchical
  multiple system.
\end{abstract}

\keywords{stars: formation --- binaries: general --- astrometry ---
radio continuum: stars --- stars: individual (IRAS~16293--2422)}

\section{Introduction}

Although the formation of stars in multiple systems is known to be a
major channel of star-formation (e.g.\ Duch\^ene et al.\ 2007 for a
recent review), our understanding of the way multiple stellar systems
form remains comparatively poorer than our comprehension of isolated
star-formation. To improve this situation, it is necessary to identify
and characterize multiple systems in the earliest phases of their
evolution (preferably during the Class 0 and I stages). Progress has
been slow, unfortunately, because very few existing instruments have
enough sensitivity and angular resolution to detect and resolve very
embedded systems even in the nearest star-forming regions. Moreover,
young stars in multiple systems are surrounded by circumstellar and/or
circumbinary disks, and often drive powerful, episodic jets that must
be correctly interpreted before a given system can be properly
characterized. As a consequence, the number of very young systems for
which the binarity is clearly established, and the system parameters
are well measured, is extremely limited. Arguably one of the most
promising cases is that of IRAS~16293--2422 (e.g.\ Wootten 1989; Mundy
et al.\ 1992; Ceccarelli et al.\ 2000; Chandler et al.\ 2005).

IRAS 16293--2422 is a well-studied very young low-mass protostellar
system located in Lynds 1689N, a dark cloud in the Ophiuchus
star-forming complex at $d$ = 120 pc (Loinard et al.\ 2008). It has a
total luminosity of about 15 \Lsun\ and has never been detected
shortward of $\lambda$ = 12 $\mu$m (e.g.\ Ceccarelli et al.\ 1998, J{\o}rgensen 
et al.\ 2008). These characteristics make IRAS~16293--2422 a {\it bona fide} Class 0
source with an age of only a few 10$^4$ yr (Andr\'e et al.\ 2000).  It
has been suspected to be multiple since Wootten (1989) and Mundy et
al.\ (1992) found it to be double both at millimeter and centimeter
wavelengths, and Mizuno et al.\ (1990) discovered that it powered a
multi-lobe outflow system. More recent observations (Hirano et al.\
2001; Castets et al.\ 2001; Chandler et al.\ 2005) confirmed these
early findings, and it is now well-established that IRAS 16293--2422
is indeed a very young multiple system. 

In all centimeter observations with an angular resolution better than
\msec{0}{3} obtained before 2006, IRAS 16293--2422 comprised three
radio sources called A1, A2 and B (Wootten 1989, Loinard 2002,
Chandler et al.\ 2005 --see Fig.\ 1). Components A1 and A2 are located
to the south-east of the system and are separated from each other by
about \msec{0}{34}, whereas component B is located about 5$''$ to the
north-west of the A1/A2 pair (Fig.\ 1). Using archival VLA
observations, Loinard (2002) and Chandler et al.\ (2005) have shown
that the position angle between A2 and A1 has increased roughly
linearly from about 45$^\circ$ in the late 1980s to about 80$^\circ$
in 2003--2005. During that same timespan, the separation between the
two sources has remained constant at about \msec{0}{34}. Two different
interpretations have been proposed for this relative motion (see
Loinard et al.\ 2007 for details). According to the first one, A1 and
A2 are two protostellar sources in a nearly circular Keplerian orbit,
seen almost exactly face-on. In the alternative possibility, A1 is
interpreted as a shock resulting from the impact of a strongly
precessing (or wobbling) jet driven by a third --as-yet undetected--
protostar in the system, presumably a companion of A2\footnote{The jet
driven by A2 itself is not aligned with the current position of A1, so
if A1 is indeed a shock feature, it must be driven by a different
source (Chandler et al.\ 2005; Loinard et al.\ 2007).}. Of course, in
the former case, the position angle between A2 and A1 should keep
increasing indefinitely, whereas in the latter, the change of position
angle with time should decelerate, and eventually reverse its course
because the jet must oscillate around an equilibrium position.

In a recent 1.3 cm image, IRAS~16293--2422 was unexpectedly found to
comprise four radio sources rather than the usual three (Fig.\ 1c
--Loinard et al.\ 2007). While components B and A1 were at the
expected positions and had the expected morphologies, component A2
appeared to have split into two sub-condensations dubbed A2$\alpha$
and A2$\beta$. Since the line joining A2$\alpha$ to A2$\beta$ was at a
position angle of about 62$^\circ$, very similar to the direction of
both the large-scale flow and the thermal jet known to be driven by
A2, Loinard et al.\ (2007) argued that A2$\alpha$ and A2$\beta$ traced
a recent bipolar ejection from A2. If this is the case, then
A2$\alpha$ and A2$\beta$ should move symmetrically away from A2, along
the direction of the jet, at velocities typical of the winds driven by
low-mass protostars (tens to hundreds of km s$^{-1}$, depending on the 
inclination of the jet). This could be
easily tested with new observations. 

As the previous discussion shows, the very nature of some of the sources 
associated with IRAS 16293--2422 remains unknown, and the exact number 
of protostars contained in the system is still uncertain. In this article, we will
present and analyze new high-resolution, high-sensitivity, radio continuum 
observation that will be used to further investigate the structure of IRAS~16293--2422.

\section{Observations} 

Two new 3.6 cm observations of IRAS~16293--2422 were obtained on 2007, August 
14 (2007.62), and 2008, December 13 (2008.95) with the {Very Large Array} (VLA) of 
the {National Radio Astronomy Observatory} (NRAO) in its most extended (A)
configuration. The
standard 3.6 cm continuum frequency setup was used, and both circular
polarizations were recorded simultaneously. The absolute flux density
was set using observations of 3C 286. For improved flux accuracy, we
did not assume 3C 286 to be a point source, but instead used a model
image provided by NRAO. The phase calibrator was PKS J1625--2527 whose
absolute position is expected to be accurate to about 2
milli-arcseconds. The data were collected during the VLA/EVLA
transition period, so the array consisted of a mixture of ``old'' VLA
antennas, and of antennas already equipped with new electronics.
The non-matched bandpass shapes between VLA and EVLA antennas 
produced significant closure errors on VLA/EVLA baselines. To correct these 
errors, we measured baseline-dependent gains using the observations of the 
phase calibrator.  The images obtained after applying these baseline-dependent 
gains are almost identical to those produced by simply flagging the VLA/EVLA
baselines.  

To optimize the angular resolution, the calibrated visibilities were
imaged using uniform weighting (ROBUST parameter set to --5),
resulting in synthesized beams of \msec{0}{35} -- \msec{0}{45}
and  $\sim$ \msec{0}{17} in the north-south and east-west
directions, respectively (see Table 1). The r.m.s.\ noise levels in the final images 
are 45--50 $\mu$Jy beam$^{-1}$. In the following sections, these new 
observations will be compared to a similar 3.6 cm observation obtained
in 2003.65, and to 0.7 cm and 1.3 cm images obtained in 2005.20 and 2006.11,
respectively  (see Fig.\ 1 --all three images were published previously by Loinard 
et al.\ 2007). The characteristics of these images are listed in Table 1
together with those of the new 3.6 cm images published here. The flux
of source B is known to be fairly constant with time at  any given wavelength
(Chandler et al.\ 2005), so measurements of the total flux of component B
provides a valuable self-consistency check on the overall calibration. The
flux found for component B in the three 3.6 cm observations included
in the present work are reported in the last column of Table 1; they are 
indeed fully consistent with each other, and with previously published figures
(Chandler et al.\ 2005).

Observations at $\lambda$ = 0.7 and 1.3 cm, designed to measure (in combination
with the 3.6 cm data presented here) the spectral index of A2$\alpha$ and A2$\beta$ 
were requested and approved both in the 2007.62 and 2008.95 runs. They were 
actually collected in the former run, but could not be properly calibrated because of poor 
weather conditions. Because of scheduling limitations, no multi-wavelengths 
data could be obtained in the 2008.95 run.

\section{Nature and origin of A2$\alpha$ and A2$\beta$}

\subsection{Structure and astrometry}

In the two new 3.6 cm observations, component B is at its expected position,
with its usual morphology (Fig.\ 1) and flux (Sect.\ 2). The structure of component
A, however, has changed significantly since early 2006 --when the last VLA
observation prior to those presented here was obtained (Fig.\ 1c). In the 
observation obtained in mid-2007 (Fig.\ 1d), component A of IRAS~16293--2422 
appears to contain four radio sources (see particularly the zoom on this region 
shown in Fig.\ 2). Although somewhat blended with the rest of the emission, 
A1 is still clearly discernable to the east of the system (Fig.\ 2).  A
radio source is also visible again at the expected position of
A2. Note that this was not the case in the 2006.11 observation, where
the emission associated with A2 was blended with
that of A2$\alpha$. The two additional sources in the system (indeed,
the two brightest ones) are located on each side of A2, and we
identify them with A2$\alpha$ and A2$\beta$. They are clearly not at
the same positions as in the 2006.11 observation. Instead, they have
moved away from A2 in a roughly symmetrical manner, A2$\alpha$ towards
the south-west, and A2$\beta$ towards the north-east.
In the observation obtained at the end of 2008, component A is (again, but 
fortuitously) composed of three sources (Figs.\ 1e and 2). Source A2 is still clearly 
identified, while source A2$\alpha$ has moved further from A2 towards the south-west. 
Component A2$\beta$ has also kept moving away from A2 (towards the 
north-east), but now appears to be blended with A1 (Fig.\ 2). 

To further investigate the nature and properties of A2$\alpha$ and 
A2$\beta$, it is interesting to compare their relative positions at the
various epochs. In 2006.11, A2$\alpha$ and A2$\beta$ were separated 
by \msec{0}{166} $\pm$ \msec{0}{003} at a position angle of 62$^\circ$ $\pm$
2$^\circ$. In the 2007.62 image, however, the separation has increased to
\msec{0}{455} $\pm$ \msec{0}{011}, but the position angle has not
changed significantly: it is now measured to be 61$^\circ$ $\pm$
3$^\circ$. From the change in their separation, we can estimate the
velocity at which A2$\alpha$ and A2$\beta$ are moving away from one
another to be 109 $\pm$ 4 km s$^{-1}$.  The errors quoted here and
later in the paper only account for the positional uncertainties of the
ejecta, they do not include the errors on the distance $d$ to the source. 
If the velocity since ejection has been constant at the value estimated above, 
the ejection must have occured 0.87 $\pm$ 0.04 yr before the 2006.11 
observation (or 2.38 $\pm$ 0.11 yr before the 2007.62 observation), i.e.\ 
in 2005.24 $\pm$ 0.04. This is --as expected-- between the epochs of the 
0.7 and 1.3 cm observations shown in Figs.\ 1 and 2, but only very shortly
 after the 0.7 cm data were gathered.

Since A2 and A2$\alpha$ are well separated in the last two 3.6 cm
observations, one can also consider the evolution of the relative 
position between these two sources. In 2007.62. they were separated
by \msec{0}{300} $\pm$ \msec{0}{015} at a position angle of 49$^\circ$ 
$\pm$ 4$^\circ$, whereas in the 2008.95 observation, the separation was
\msec{0}{494} $\pm$ \msec{0}{014} and the position angle 61$^\circ$ 
$\pm$ 3$^\circ$. Thus, A2$\alpha$ appears to be moving away from A2
at 82 $\pm$ 9 km s$^{-1}$. Assuming again that the velocity has not changed
appreciably, the ejection must have occurred 3.41 $\pm$ 0.37 yr before the
2008.95 observation, i.e.\ in 2005.54 $\pm$ 0.37. This is in good agreement
with the date estimated above from the relative motion between A2$\alpha$
and A2$\beta$, suggesting that the assumption of constant velocity is reasonable. 
Note that the relative velocity between A2 and A2$\alpha$ derived 
here is somewhat larger than half of the velocity 
between A2$\alpha$ and A2$\beta$ calculated above. This shows 
that A2$\alpha$ is moving away from A2 somewhat faster than 
A2$\beta$\footnote{One could argue that this result might also be 
consistent with an acceleration of the velocity of the ejecta since the 
relative velocity between A2$\alpha$ and A2 is based on more
recent observations that the estimate of the relative velocity between 
A2$\alpha$ and A2$\beta$. We favor the interpretation given in the text
because the two recent 3.6 cm observations (Fig.\ 2) clearly show that 
A2$\alpha$ has moved farther from A2 than A2$\beta$.}. Indeed, one can 
estimate the relative velocity between A2$\beta$ and A2 combining
the present result and the relative velocity between A2$\alpha$ and 
A2$\beta$ calculated earlier. We obtained (assuming again constant
velocities) 26 $\pm$ 10 km s$^{-1}$. Thus A2$\beta$ appears to move
away from A2 about three times faster than A2$\alpha$. Bipolar ejections 
usually produce somewhat more symmetric patterns. It should be mentioned,
however, that the north-east and south-west lobes of the molecular
outflow driven by A2 have long been known to be very asymmetric. 
For instance, SiO emission is very strong in the direction of the north-east 
lobe and nearly absent towards the south-west counterpart (e.g.\ Castets 
et al.\ 2001; Hirano et al.\ 2001). Since SiO is a good tracer of shocks
between jets and circumstellar material, this most likely indicates that the
region to the south-west of component A contains relatively little dense 
gas capable of decelerating the ejecta.

In summary, A2$\alpha$ and A2$\beta$ appear to behave kinematically 
exactly as would be expected if they were ejecta from A2: they are moving 
(in projection) at 30--80 km s$^{-1}$ away from A2 along the direction (P.A.\ $\sim$ 
60$^\circ$) of the outflow known to be powered by A2. The true velocity of the jet 
must be of the order of the escape velocity 
from A2. As we will see in Sect.\ 4, A2 is likely to be a $\sim$ 1.5 \Msun\ protostar. 
The radius at which jets are launched is usually believed to be a few stellar radii 
($\sim$ 3 $R_*$), and very young stars are a few times larger than their main 
sequence counterparts of the same mass. It is, therefore, reasonable to assume 
that the radius of the protostar associated with A2 is about 3 \Rsun, and that the 
escape velocity should be calculated at $\sim$ 10 \Rsun. Under these assumptions, 
we obtain $V_{esc}$ $\approx$ 240 km s$^{-1}$. To obtain a rough estimate of the 
orientation of the jet, we assume that this value provides a reasonable estimate of 
the true current velocity of the jet (this would require, in particular, that the jet has 
suffered little deceleration since it was launched).  The projected velocity of the 
ejecta in only 30--80 km s$^{-1}$, so the jet powered by A2 must be oriented along 
a direction only 10$^\circ$--15$^\circ$ from the line of sight. We conclude that A2 
drives a flow oriented almost along the line of sight, and that A2$\alpha$ and A2$\beta$ 
are ejecta along that flow. Episodic bipolar mass ejections are known to occur in 
young stars (e.g.\ Marti et al.\ 1995). To our knowledge, this is the first time, however, 
that an ejection is actually observed from the very beginning: we seem to have
witnessed the very birth of a Herbig-Haro pair.

\subsection{Properties of the ejecta}

The centimeter emission produced by winds and ejecta from low-mass 
stars is thought to be nearly entirely of free-free origin (e.g.\ Anglada 1995, Shang et al.\
2004). 
As detached clumps, A2$\alpha$ and A2$\beta$ are likely less dense than 
the so-called thermal jets associated with the central regions of winds driven 
by young stars. As a consequence, the free-free emission from A2$\alpha$
and A2$\beta$ is likely to be optically thin. In the absence of simultaneous
multi-frequency observations (see Sect.\ 2), it is somewhat 
hazardous to estimate their spectral index and ascertain the characteristics of the
emission. We note, however, that our data are fully consistent with optically thin 
free-free emission. The source A2$\beta$ was well-resolved in the 2006.11 1.3 cm
data and in the 2007.62 3.6 cm observations. The spectral index derived from
these two observations is $\alpha$ = --0.09 $\pm$  0.05, in excellent agreement
with the expected value ($\alpha$ = --0.1) for optically thin free-free emission.

To further constrain the properties of the ejecta, we will concentrate on A2$\alpha$, 
because it is well-resolved from the other sources in both of our 3.6 cm data sets. 
Within the errors, the 3.6 cm flux of A2$\alpha$ does not appear to have changed
much between the two observations (0.62 $\pm$ 0.09 mJy in 2007.62
and 0.93 $\pm$ 0.10 in 2008.95 --Table 2). The angular size of the emission
(deconvolved from the primary beam) was found to be \msec{0}{21} $\times$ 
\msec{0}{08} in the 2008.95 data, whereas the emission was only resolved in 
one direction in the 2007.62 observations. In the resolved dimension, the
angular size was \msec{0}{14}, whereas in the other direction, the emission 
came from a region smaller than \msec{0}{17}. Thus, the mean angular size 
of the emission was about \msec{0}{14} in 2008.95 and less than about 
\msec{0}{15} in 2007.62.

Assuming optically thin free-free emission, the mass of ionized gas can
be calculated from the radio flux as (e.g.\ Rodr\'{\i}guez et al.\ 1980):

\begin{equation}
{M_i \over M_\odot} = 3.39 \times 10^{-5} \left( {S_\nu \over  1~\mbox{mJy}} \right)^{0.5} \left( {\nu \over 1~\mbox{GHz}} \right)^{0.05} \left( {T_e \over 10^4~\mbox{K}} \right)^{0.175} \left( {\theta \over 1''} \right)^{1.5} \left( {d \ \over 1~\mbox{kpc}} \right)^{2.5}.
\end{equation}

\noindent
Using the numbers above and $T_e$ = 10$^4$ K, we obtain, for A2$\alpha$, 
$M_i$ $\approx$ (1.00 $\pm$ 0.05) $\times$ 10$^{-8}$ \Msun\ using the 2008.95 
observations (once again, the quoted uncertainty does not include the
errors on the distance to the source), and $M_i$ $\lesssim$ 0.9 $\times$ 10$^{-8}$ \Msun\ using 
the 2007.62 data. Assuming that both ejecta have similar masses,
the bipolar ejection event reported here corresponds to a total mass
of about 2 $\times$ 10$^{-8}$ \Msun.

From the observed radio flux, one can also calculate the electron density
$n_e$ of the ejecta (e.g.\ Rodr\'{\i}guez et al.\ 1980):

\begin{equation}
{n_e \over 1~\mbox{cm}^{-3}} = 7.8 \times 10^{3} \left( {S_\nu \over  1~\mbox{mJy}} \right)^{0.5} \left( {\nu \over 1~\mbox{GHz}} \right)^{0.05} \left( {T_e \over 10^4~\mbox{K}} \right)^{0.175} \left( {\theta \over 1''} \right)^{-1.5} \left( {d \ \over 1~\mbox{kpc}} \right)^{-0.5}.
\end{equation}

\noindent With the observed parameters of the emission, we get 
$n_e$ = 4.4 $\times$ 10$^{5}$ cm$^{-3}$ for A2$\alpha$. Interestingly, the 
recombination timescale at that density is about 6 months. Since the ejecta 
have remained ionized at least since 2006.11 (when they were first detected) 
and most certainly since their creation around 2005.3 (see above),  some 
mechanism must provide energy to keep them ionized. The most likely candidates 
are shocks either with the surrounding medium or internal to the jets. 
To remain ionized, the ejecta require an ionization rate of $\sim 10^{42}$ 
s$^{-1}$. Assuming that 13.6 eV of energy are required ionization, a power of 
$\sim 2 \times 10^{31}$  erg s$^{-1}$ is needed. If this power is produced by 
the kinetic energy of the ejecta, we expect that the ejecta should decelerate
at a rate of about 12.5 km s$^{-1}$ yr$^{-1}$ (for an initial velocity of 240 km 
s$^{-1}$ and a mass of $10^{-8}$ $M_\odot$). The true deceleration would
be somewhat smaller if the ejecta were only partially ionized as seems to be
commonly the case  (e.g.\ Podio et al.\ 2009). In any case, since the jet appears 
to be only 10--15$^\circ$ from the line of sight, the projected deceleration would
only be 2--3 km s$^{-1}$ yr$^{-1}$ and would be undetectable with the existing 
observations.

\subsection{Origin of the ejecta}

Detached clumps have long been known to exist in the jets driven by 
young stars. They can be created in (at least) two ways. One possibility 
is if the driving source experiences episodic increases in its 
mass loss rate. This hypothesis is often the preferred one to explain the 
presence of symmetric pairs of well-defined HH knots in 
evolved outflows (e.g.\ Arce \& Goodman 2002). An alternative possibility
is if the mass loss rate remains constant, but the ejection velocity increases
abruptly from an initial value $v_i$ to a final (larger) one $v_f$. In this situation, 
a working surface where material accumulates is created 
at the interface between the two winds. This naturally creates a gas condensation 
which can eventually become a detached clump (Masciadri \& Raga 2001).
Both of these mechanisms are plausible scenarios for the creation of
A2$\alpha$ and A2$\beta$, and it would be interesting to be able to distinguish 
between them. 
 
Ejection and accretion are believed to be intimately linked in protostars, 
so if A2$\alpha$ and A2$\beta$ were created during an episode of
increased mass loss, one would expect that increased accretion would
also have been occuring. Assuming that the mass ejection rate is about 10 times smaller than the 
accretion rate (e.g.\ Hartmann \& Kenyon 1996), then the total mass accreted during this
episode must have been about 2 $\times$ $10^{-7}$ \Msun. Very young
protostars derive much of their luminosity from accretion, so one could
wonder if such an episode of increased accretion might have produced a
detectable increase in the total luminosity of IRAS~16293--2422. To
try and answer that question, one must characterize the timescale of
the ejection/accretion episode.

From Fig.\ 2, it is clear that by 2007.62, the ejecta had become detached 
from A2. As a consequence, a conservative upper limit on the timescale
of the ejection event is 2.34 yr, the time elapsed between the ejection and
the 2007.62 observation (Sect.\ 3.1). The corresponding lower limit on the 
accretion rate during the event would be $\dot{M}_{min}$ $\approx$ 8.5 
$\times$ 10$^{-8}$ \Msun\ yr$^{-1}$, and the corresponding lower limit on 
the excess accretion luminosity $L_{acc, min}$ $\approx$ 1.3 \Lsun\ (here
we have assumed that all the accretion energy is released on the stellar 
surface, and that the stellar radius is 3 \Rsun). The bolometric luminosity
of IRAS~16293--2422 is about 15 \Lsun\ (Sect.\ 1), so the increased  accretion
should have produced a $\sim$ 10\% increase in the total luminosity of the
source during the assumed 2.34 yr duration of the event. 

Of course, the duration of the event might have been significantly shorter
than 2.34 yr. An alternative way of estimating the characteristic timescale
is the following. The ejecta are currently about \msec{0}{14}
($\equiv$ 17 AU) across (they were likely smaller in the past since A2$\beta$
appeared to be unresolved in the 1.3 cm observations obtained in 2006.11).
A clump of that size moving at $\sim$ 240 km s$^{-1}$ would become fully 
detached from its ejecting star in about 1.06 $\times$ 10$^{7}$ s ($\equiv$ 0.34 yr, just about
4 months). If this provides a good estimate of the true duration of the
ejection event, then the mass accretion rate during the event would have been
$\dot{M}$ $\approx$ 6 $\times$ 10$^{-7}$ \Msun\ yr$^{-1}$, and 
the corresponding excess accretion luminosity $L_{acc}$ $\approx$ 9 \Lsun.
This would have produced a 60\% increase in the total luminosity of the source
during the 4 month duration of the event.

We conclude that if the creation of the A2$\alpha$/A2$\beta$ pair is the result
of an increase in the mass accretion/ejection rates of component A2, their birth
should have been accompanied by a 10--60\% increase in the total luminosity 
of IRAS~16293--2422. Most of the luminosity of IRAS~16293--2422 is radiated 
in the far-infrared and sub-millimeter parts of the electromagnetic spectrum, so 
the corresponding luminosity increase should be sought there. In particular, 
there have been several sub-millimeter observations of IRAS~16293--2422 obtained 
with the Sub-Millimeter Array (SMA) in the last few years (T.\ Bourke, private 
communication). Note that IRAS~16293--2422 is a multiple system (Sect.\ 1) and that the only
member of the system which should have experienced an increase in luminosity
is component A2. While components A1 and A2 are not resolved in SMA
observations, component A is well resolved from component B (e.g.\ Chandler
et al.\ 2005). Moreover, component A and B happen to have similar sub-millimeter 
fluxes (e.g.\ Chandler et al.\ 2005). As a consequence, the total sub-millimeter 
luminosity increase for component A in SMA images ought to be at least 20--120\% 
and should be detectable. 

We note that an ejection/accretion event such as the one considered in
the present discussion would remain fairly modest in comparison with the more 
spectacular FUOr events which produce increases in the luminosity, mass 
accretion rates, and mass ejection rates of several orders of magnitude (Hartmann \& 
Kenyon 1996). In this sense, we would only have witnessed a ``mini-outburst''. 
The very fact that such a mini-outburst would have been observed, 
however, would indicate that they most likely occurred quite frequently. They may, therefore, 
offer more tractable evidence of the relation between accretion and outflow 
phenomena than the more dramatic, but much less common FUOr events.

If the creation of the A2$\alpha$/A2$\beta$ pair is related to a modification of
the jet ejection speed without an associated increase in mass accretion rate, 
no change in the luminosity of IRAS~16293--2422 is expected. Thus the
presence or absence of an associated luminosity increase would be a good
discriminant between the two possibilities. We note that
a $\sim$ 30\% increase in the velocity of a fairly modest underlying jet (with
$\dot{M}$ of a few 10$^{-7}$ \Msun\ yr$^{-1}$) lasting for a few months, would be
sufficient to accumulate a mass of order 10$^{-8}$ \Msun\ in a working surface.

\section{Relative motion between A1 and A2}

The motion of A1 relative to A2 between the late 1980s and 2005 has
been investigated in detail by Loinard (2002) and Chandler et al.\
(2005). Neither the 1.3 cm observation obtained in 2006.11 (where A2 
is blended with A2$\alpha$) nor the 2008.95 data (where A1 is blended
with A2$\beta$) can be used to track the relative motion between
A1 and A2 further. However, A1 and A2 are well resolved in the
2007.62 observation. The value of their relative position angle in
this new observation is 90 $\pm$ 3$^\circ$, significantly larger than
the position angle in 2003--2005 (Chandler et al.\ 2005, Loinard et
al.\ 2007 -- Fig 3). It is, however, in good agreement with the
general evolution of the position angle since the late 1980s. The
separation between A1 and A2 in the 2007.62 3.6 cm observation is
0.365 $\pm$ 0.010 in good agreement with all previous
measurements (Fig.\ 3).

Thus, the position angle between A1 and A2 has now changed by
more than 40$^\circ$ since the late 1980s (from less than 50$^{\circ}$ then, to
90$^{\circ}$ now). Moreover, there is no indication in the data
that the rate of change is, in any way, decelerating. Although further
monitoring will be needed in the coming years and decades, these
characteristics are already difficult to reconcile with the idea of a
precessing jet. Indeed, the maximum precession angles typically
observed in low-mass young stars are less than 10$^\circ$, and rarely
exceed 15$^\circ$ (e.g.\ Matthews et al.\ 2006). The relative motions
between A2 and A1 are more readily explained in terms of a Keplerian
orbit between two protostars. The fit shown in Fig.\ 3b implies a rate
of position angle change with time of 1.98$^\circ$ yr$^{-1}$,
corresponding to an orbital period of 182 yr.  For a circular orbit in
the plane of the sky\footnote{The fact that the jet from A2 is likely
nearly along the line of sight (Sect.\ 3.1) would be consistent with
an A1/A2 orbit nearly in the plane of the sky; in young binary
systems, a coplanarity between the orbital plane and the orientation
of the disks corresponds to the most stable configuration.}
(which would produce the observed linear increase
of the position angle), the total mass of the A1+A2 system would be
almost exactly 2 \Msun\ (assuming a separation of \msec{0}{34} and a
distance to Ophiuchus of 120 pc --Loinard et al.\ 2008). Based on an analysis of the
absolute proper motions, Loinard (2002) and Chandler et al.\ (2005)
have argued that the center of mass of the A1/A2 system must be
significantly closer to A2 than to A1 in this Keplerian scheme,
implying that A2 must be significantly more massive than A1.  A
reasonable estimate would be 1.5 \Msun\ for A2 and 0.5 \Msun\ for A1. 
This would be in reasonable agreement with the bolometric luminosity of
IRAS~16293--2422 ($\sim$ 15 \Lsun).

In a recent submillimeter observation, yet another compact source was
detected toward component A (Chandler et al.\ 2005). This object
(called Ab) is located about \msec{0}{64} to the northeast of the
A1/A2 pair (see Fig.\ 1). Since it has only been detected at one
wavelength so far, the nature of Ab is difficult to assess. Chandler
et al.\ (2005) argued that it might be another protostellar source in
the system, but the lack of a strong compact counterpart at 0.7 cm in
recent VLA data (Loinard et al.\ 2007) might favor an interpretation
in terms of a starless clump. If Ab were a protostar, then component A
would overall be (at least) a triple system, whereas it might only be
double if Ab is a starless clump. In any case, taking into account the
fact that component B --to the north-west of the system-- is also
associated with a protostar (e.g.\ Rodr\'{\i}guez et al.\ 2005;
Chandler et al.\ 2005) and that components A and B are located in the
center of a common, dense, centrally condensed, envelope (e.g.\ Looney
et al. 2003), we must conclude that IRAS~16293--2422 is most likely a
very young hierarchical multiple system.

IRAS~16293--2422 has long been known to drive a multi-lobe outflow
system composed of two compact bipolar flows at P.A.\ $\sim$
60$^\circ$ and $\sim$ 110$^\circ$, and a larger monopolar lobe located
farther east (Mizuno et al.\ 1990). Recent high-resolution
SMA observations of the two compact outflows
strongly suggest that both are driven from within the A component of
IRAS~16293-2422 (Yeh et al.\ 2008). As mentioned earlier, the outflow
at P.A.\ $\sim$ 60$^\circ$ is now known to be driven by A2. Since the
A1/A2 pair is most likely a compact binary system, it is tempting to
associate the flow at P.A.\ $\sim$ 110$^\circ$ with A1.
Interestingly, while the dynamical ages of the two flows are
comparable (500 to 1,000 yr), the mechanical luminosity of the flow at
P.A.\ $\sim$ 60$^\circ$ (from A2) is about twice that of the flow at
P.A.\ $\sim$ 110$^\circ$ (that we tentatively attribute to A1). This
would be consistent with the higher mass of A2 as compared to A1.

\section{Conclusions and perspectives}

In this article, we presented two new high-quality 3.6 cm images of
the young protostellar system IRAS~16293--2422 obtained in August 2007
and December 2008 with the Very Large Array. These observations confirm 
that the radio sources A2$\alpha$ and A2$\beta$ recently identified in the system are
ejecta from the protostar A2, and that we seem to have witnessed the
very birth of a pair of Herbig-Haro knots. The mass of each of the ejecta
is estimated to be $\sim$ 10$^{-8}$ \Msun. If the creation of the ejecta
was related to an increase in mass accretion rate, the birth of 
A2$\alpha$/A2$\beta$ must have been accompanied by an increase in
the total luminosity of IRAS~16293--2422 of 10--60\%.

Source A2 itself, which was blended with A2$\alpha$ in recent
observations, is again visible in the data. This allows us to further monitor the
relative motion between A1 and A2, and to provide very suggestive
evidence that the A1/A2 pair is a tight binary system. Including
component B to the north-west of the system, IRAS~16293--2422,
therefore, appears to be a very young hierarchical multiple
system. Observations similar to those presented here obtained in
the coming few years to decades ought to provide a very accurate determination
of the mass of the various protostars in IRAS~16293--2422, and a very detailed
characterization of this nearby very young multiple stellar system.

\acknowledgements L.L., L.F.R., and P.A.\ acknowledge the
financial support of DGAPA, UNAM and CONACyT, M\'exico.
D.J.W. acknowledges partial support from NASA Origins of Solar Systems
Program Grant NAG5-11777. NRAO is a facility of the National Science
Foundation operated under cooperative agreement by Associated
Universities, Inc.

\clearpage

\begin{deluxetable}{lccccccc}
\tablecaption{Observations}
\tablehead{
\colhead{Epoch}       &  
\colhead{Project}      &
\colhead{$\nu$}         &
\colhead{Synthesized Beam} &
\colhead{$\sigma$} &
\colhead{$F_\nu(B)$\tablenotemark{a}} \\
& & (GHz) & & ($\mu$Jy beam$^{-1}$) & (mJy)}
\startdata
2003.65 & AL589 & 8.46    & \sbeamp{0}{39}{0}{19}{6.8} & 36 & 0.61 $\pm$ 0.08\\%
2005.20 & AC778 & 43.23 & \sbeamm{0}{30}{0}{17}{1.9} &179 & 26.5 $\pm$ 0.5\\%
2006.11 & AL672 & 22.46 & \sbeamm{0}{13}{0}{06}{0.5} & 47 & 6.5 $\pm$ 0.2\\%
2007.62 & AL703 & 8.46   & \sbeamm{0}{35}{0}{16}{0.9} & 45 & 0.61 $\pm$ 0.09\\%
2008.95 & AL728 & 8.46   & \sbeamm{0}{45}{0}{18}{24.6} & 52 & 0.60 $\pm$ 0.10\\%
\enddata
\tablenotetext{a}{Flux density of component B.}
\end{deluxetable}

\begin{deluxetable}{lccccccc}
\tablecaption{Source positions and fluxes}
\tablehead{
\colhead{Epoch}       &  
\colhead{$\nu$}         &
\colhead{Source} &
\colhead{$\alpha$ (J2000.0)} &
\colhead{$\delta$ (J2000.0)} &
\colhead{$F_\nu$}\\
& (GHz) & & 16$^h$32$^m$ & --24$^\circ$28$'$ & (mJy)}
\startdata
2003.65 & 8.46    & A1                       & \mmsec{22}{8823} $\pm$ \mmsec{0}{0004} & \msec{36}{301} $\pm$ \msec{0}{011} & 0.81 $\pm$ 0.08\\%
                &             & A2                       & \mmsec{22}{8544} $\pm$ \mmsec{0}{0002} & \msec{36}{348} $\pm$ \msec{0}{007} & 1.35 $\pm$ 0.08\\%
\\[-0.3cm]
2006.11 & 22.46  & A1                      &  \mmsec{22}{8829} $\pm$ \mmsec{0}{0002} & \msec{36}{373} $\pm$ \msec{0}{004} & 0.78 $\pm$ 0.09\\%
                 &              & A2$\beta$        &  \mmsec{22}{8527} $\pm$ \mmsec{0}{0001} & \msec{36}{438} $\pm$ \msec{0}{004} & 0.64 $\pm$ 0.05\\%
                   &            & A2+A2$\alpha$ &  \mmsec{22}{8594} $\pm$ \mmsec{0}{0006} & \msec{36}{398} $\pm$ \msec{0}{006} & 3.58 $\pm$ 0.27\\%
\\[-0.3cm]
2007.62 & 8.46    & A1                       & \mmsec{22}{8799} $\pm$ \mmsec{0}{0006} & \msec{36}{393} $\pm$ \msec{0}{013} & 0.50 $\pm$ 0.08\\%
                &             & A2$\beta$         & \mmsec{22}{8657} $\pm$ \mmsec{0}{0003} & \msec{36}{371} $\pm$ \msec{0}{009} & 0.70 $\pm$ 0.07\\%
                &             & A2                       & \mmsec{22}{8531} $\pm$ \mmsec{0}{0004} & \msec{36}{394} $\pm$ \msec{0}{011} & 0.60 $\pm$ 0.07\\%
                &             & A2$\alpha$       & \mmsec{22}{8366} $\pm$ \mmsec{0}{0006} & \msec{36}{592} $\pm$ \msec{0}{015} & 0.62 $\pm$ 0.09\\%
\\[-0.3cm]
2008.95 & 8.46    & A1+A2$\beta$  & \mmsec{22}{8803} $\pm$ \mmsec{0}{0004} & \msec{36}{414} $\pm$ \msec{0}{009} & 0.93 $\pm$ 0.09\\%
                &             & A2                       & \mmsec{22}{8548} $\pm$ \mmsec{0}{0008} & \msec{36}{481} $\pm$ \msec{0}{014} & 1.26 $\pm$ 0.13\\%
                &             & A2$\alpha$       & \mmsec{22}{8232} $\pm$ \mmsec{0}{0005} & \msec{36}{723} $\pm$ \msec{0}{011} & 0.93 $\pm$ 0.10\\%
\enddata
\end{deluxetable}

\clearpage

\begin{figure*}[!t]
\centerline{\includegraphics[width=0.72\textwidth,angle=270]{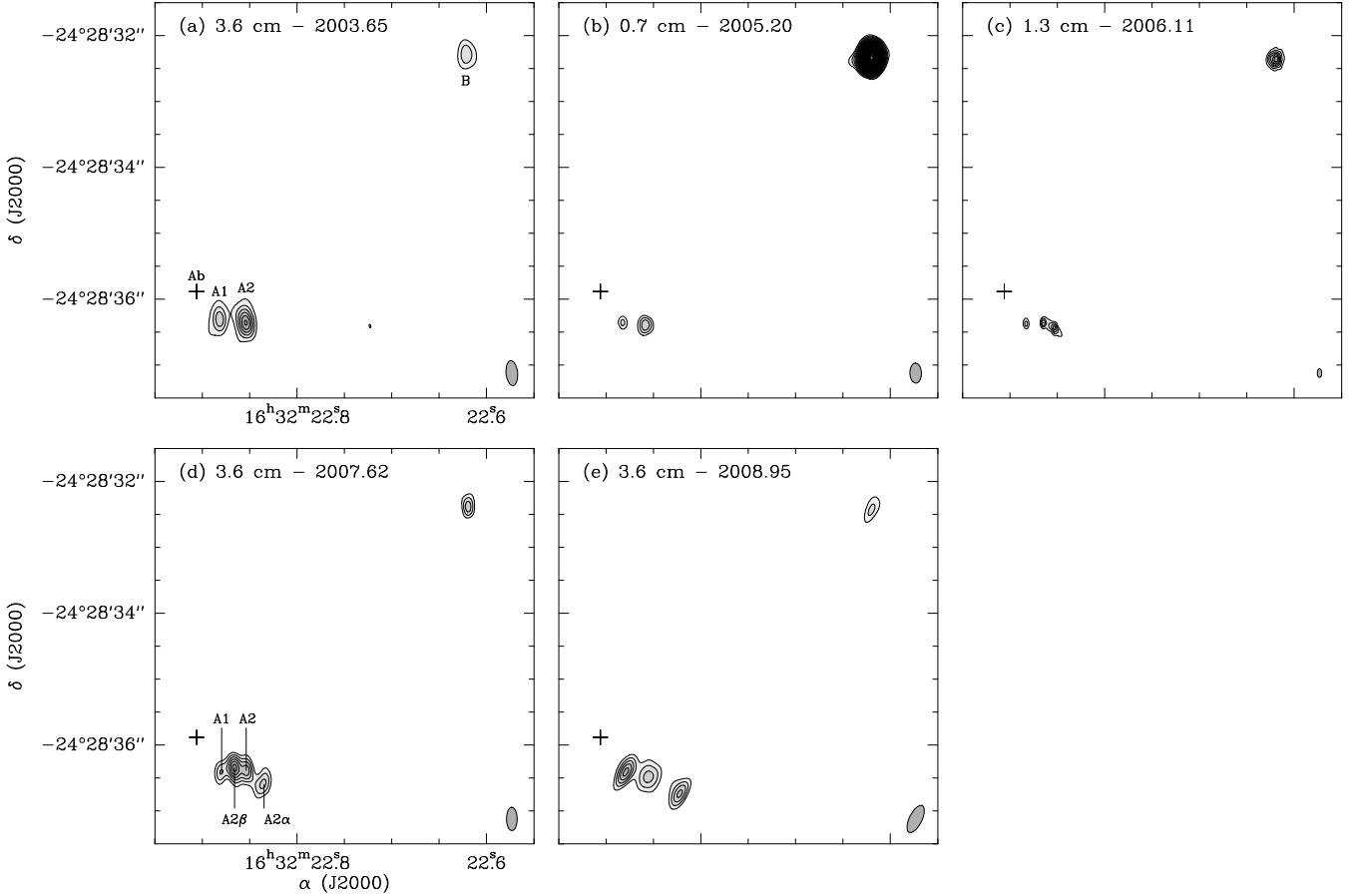}}
\caption{Comparison between five recent radio images of
IRAS~16293--2422. The synthesized beam is shown at the bottom
right of each panel and is given explicitly in Tab.\ 1. The radio sources
A1, A2, B, A2$\alpha$, and A2$\beta$ are labelled in panels (a) and
(c). The position of the sub-millimeter source Ab is shown as a cross
to the north-east of component A in each panel (see panel (a)).
(a) 3.6 cm
obtained in 2003.65. The first contour and the contour interval are at 
0.15 mJy beam$^{-1}$.  (b) 0.7 cm image obtained in 2005.20. The first
contour is at 1.5 mJy beam$^{-1}$, and the contour interval is 0.3 mJy
beam$^{-1}$.  (c) 1.3 cm image obtained in 2006.11. The first contour and the contour
interval are at 0.0255 mJy beam$^{-1}$.  (d) 3.6 cm image obtained in 2007.52. The
first contour is at 0.2 mJy beam$^{-1}$, and the contour interval is
0.1 mJy beam$^{-1}$. (e) 3.6 cm
obtained in 2008.95. The
first contour is at 0.3 mJy beam$^{-1}$, and the contour interval is
0.15 mJy beam$^{-1}$.
Note how the morphology of component A changes with time.}  
\end{figure*}

\clearpage

\begin{figure}[!t]
\centerline{\includegraphics[width=0.37\textwidth,angle=0]{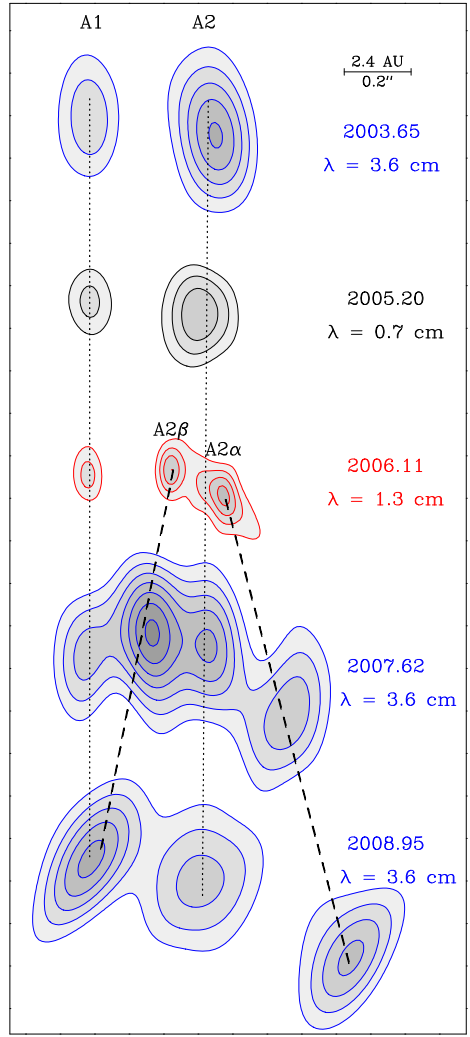}}
\caption{Zoom on the A component, from Fig.\ 1. The contours are the
same as in Fig.\ 1. The linear and angular scales are shown at the 
top right of the figure.} 
\end{figure}

\clearpage

\begin{figure*}[!t]
\centerline{\includegraphics[width=0.35\textwidth,angle=270]{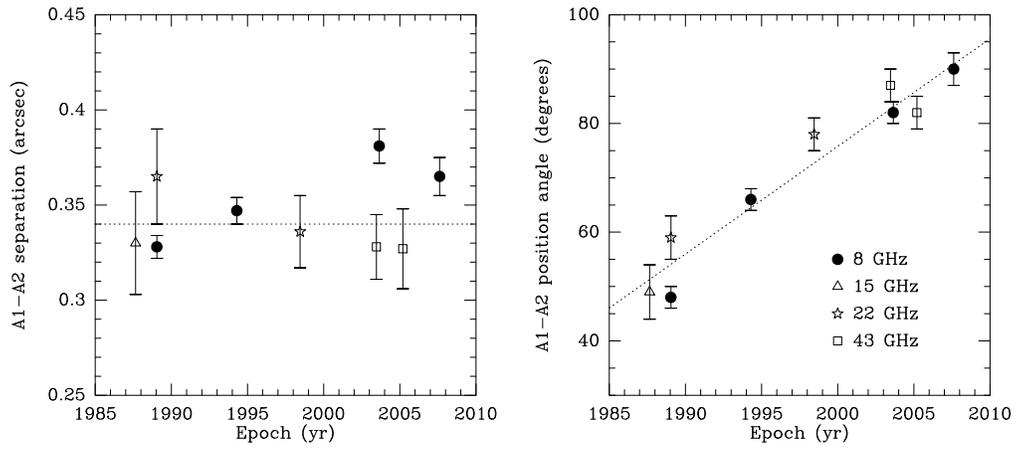}}
\caption{Time evolution of the separation (left) and position
angle (right) between A2 and A1.}  
\end{figure*}

\end{document}